# AI-based Medical e-Diagnosis for Fast and Automatic Ventricular Volume Measurement in the Patients with Normal Pressure Hydrocephalus

Xi Zhou[1,#], Qinghao Ye[2,#], Xiaolin Yang[1], Jiakun Chen[1], Haiqin Ma[1], Jun Xia[1,*], Javier Del Ser[3,†], Guang Yang[4,5,*,†]

[1] Department of Radiology, The First Affiliated Hospital of Shenzhen University, Shenzhen University, Shenzhen second people's hospital, 3002 SunGang Road West, Shenzhen, 518035, Guangdong Province, China

[2] Department of Computer Science and Engineering, University of California, San Diego, La Jolla, CA, USA

[3] University of the Basque Country (UPV/EHU), 48013 Bilbao, Spain

[4] Royal Brompton Hospital, London, UK

[5] National Heart and Lung Institute, Imperial College London, London, UK

[#] Xi Zhou and Qinghao Ye contributed equally to this work.

[†] Javier Del Ser and Guang Yang are co-last authors of this work.

[*] Corresponding authors: Jun Xia and Guang Yang.

Jun Xia

E-mail: xiajun@email.szu.edu.cn

Phone: +86 13828792422

Fax: +86 0755-83356952

Guang Yang

E-mail: g.yang@imperial.ac.uk

Phone: +44 02073528121

**Abstract**

Based on CT and MRI images acquired from normal pressure hydrocephalus (NPH) patients, using machine learning methods, we aim to establish a multi-modal and high-performance automatic ventricle segmentation method to achieve efficient and accurate automatic measurement of the ventricular volume. First, we extract the brain CT and MRI images of 143 definite NPH patients. Second, we manually label the ventricular volume (VV) and intracranial volume (ICV). Then, we use machine learning method to extract features and establish automatic ventricle segmentation model. Finally, we verify the reliability of the model and achieved automatic measurement of VV and ICV. In CT images, the Dice similarity coefficient (DSC), Intraclass Correlation Coefficient (ICC), Pearson correlation, and Bland-Altman analysis of the automatic and manual segmentation result of the VV were 0.95, 0.99, 0.99, and 4.2 ± 2.6 respectively. The results of ICV were 0.96, 0.99, 0.99, and 6.0 ± 3.8 respectively. The whole process takes 3.4 ± 0.3 seconds. In MRI images, the DSC, ICC, Pearson correlation, and Bland-Altman analysis of the automatic and manual segmentation result of the VV were 0.94, 0.99, 0.99, and 2.0 ± 0.6 respectively. The results of ICV were 0.93, 0.99, 0.99, and 7.9 ± 3.8 respectively. The whole process took 1.9 ± 0.1 seconds. We have established a multi-modal and high-performance automatic ventricle segmentation method to achieve efficient and accurate automatic measurement of the ventricular volume of NPH patients. This can help clinicians quickly and accurately understand the situation of NPH patient's ventricles.

# 1 Introduction

In 1965, Hakim and Adams [1] first proposed the concept of normal pressure hydrocephalus (NPH), that clinical symptoms are gait disorder, urinary incontinence, and dementia; the pressure of the cerebrospinal fluid during lumbar puncture is normal; the imaging manifestations are communicating hydrocephalus [2, 3]. In most parts of the world, the number of elderly and dementia patients is increasing [4]. Studies have shown that the prevalence of NPH is as high as 5.9% among the elderly over 80 [5]. As a kind of dementia disease that can be treated in the elderly [2], NPH is of increasing clinical importance [6]. On the one hand, early diagnosis and surgical treatment may increase the likelihood of a good prognosis for patients with NPH [7]. On the other hand, NPH has a spectrum of disease development, and radiological signs precede clinical symptoms [8]. Morphological evaluation of CT or MRI is essential for screening and diagnosing patients with NPH. Enlargement of the ventricle is an imaging feature of NPH [2]. More importantly, similar to NPH, enlarged ventricles are also related to cognitive and gait disorders [9]. Therefore, it is very necessary to evaluate the ventricles of patients with NPH.

In the past, researchers manually segmented the ventricle to calculate the volume. But this method needs to be based on professional knowledge [10, 11] and is very time and energy consuming [12-14]. More importantly, it is prone to human errors [15]. Therefore, it is precise because of these shortcomings that manual calculation of ventricular volume is often only used in clinical research with small samples and is difficult to apply in clinical practice with large samples [16, 17]. With the development of science and technology, artificial intelligence methods [18, 19] and the field of medical imaging are getting closer and closer [20, 21]. In recent years, some researchers have proposed using artificial intelligence methods to automatically extract ventricle features and calculate ventricular volume [22-24] through medical image analysis techniques [25, 26]. However, these methods for automatically measuring the volume of the ventricle are single mode. What's more, these methods cannot be applied to CT and MRI images at the same time, let alone calculate ventricular volume (VV) and intracranial volume (ICV) at the same time.

So, our purpose is based on CT and MRI images of NPH patients, using machine learning methods, to achieve efficient and accurate automatic measurement of the ventricular volume of NPH patients. This will lay a solid foundation for the development of a large sample of the ventricular volume of NPH patients, and it will also help clinicians quickly and accurately assess the ventricular condition of NPH patients [12, 14, 27].

## 2 Method

### 2.1 Patient and instrument information

Strictly follow the guidelines [28], we retrospectively extracted CT and MRI images of the brains of 143 definite NPH patients in Shenzhen Second People's Hospital from January 1, 2014 to December 31, 2020. Among the 143 patients with definite NPH, 38 patients had only CT images, 46 patients had only MRI images, and 84 patients had both CT images and MRI images. Therefore, we obtained the brain CT images of 122 patients (38 + 84 = 122) with definite NPH and the brain MRI images of 130 patients (46 + 84 = 130) with definite NPH. Their characteristic information is listed in Table 1. In this study, a total of three MRI equipment and two CT equipment are included. Their specific parameters are as described in Table 2. The flow chart and network structure of the whole process is shown in Figure 1.

### 2.2 Ethical approval

All procedures performed in studies involving human participants were in accordance with the ethical standards of the institutional and/or national research committee and with the 1964 Helsinki declaration and its later amendments or comparable ethical standards. This study passed the ethical approval of The First Affiliated Hospital of Shenzhen University's bioethics committee (approval no. KS20190114001), and the researchers all signed the informed consent form.

### 2.3 Manually label the ventricles and the intracranial volume

First, a radiologist with more than 5 years of clinical work experience manually marked the VV and ICV. Then, a neurosurgeon with more than 10 years of clinical work experience adjusted the manually marked results. Finally, an anatomy expert with more than 20 years of work experience reviewed the manual annotation results.

### 2.4 Computer processor information

All of the experiments were conducted on a machine with 48 cores Intel Xeon Platinum (Cooper Lake) 8369 processor and 192GB memory. For training the model, the PyTorch framework was utilized with 4 Nvidia Titan RTX GPUs.

**2.5 Image preprocessing**

In the data pre-processing stage, we first normalized images using the z-score normalization through subtracting its mean and divided by its standard deviation. For handling the anomaly pixels in scanned CT/MRI volumes, we clipped them within the range of 2.5-quantile and 97.5-quantile. Then, we augmented the data with random horizontal flipping with 0.5 probability and random scaling of hue, saturation, and brightness within the range [0.8, 1.2]. Besides, we scaled the images and masks using the bicubic interpolation method and nearest interpolation respectively.

**2.6 Machine learning process**

First, we randomly select 20% of the data from the dataset as the test set, and the rest are regarded as the training set. Then, the multi-modal segmentation model is described in the following part.

In real-world scenarios, the thick-slice images can be easily obtained since they do not require heavy human annotations, whereas annotating thin-slice images is labor-intensive. Besides, the distribution of thick-slice images and thin-slice images are distinctive, which leads to the domain shift problem that hinders the ability of deep learning models. Moreover, CT and MRI images are easy to obtain, to leverage these two modalities of images. We propose a segmentation model that can automatically segment the CT images and MRI images regardless of the thickness. In our previous research, the feasibility of this model was verified [21, 29].

Our goal is to utilize the thick images of different modalities to minimize the performance gap between thick-slice CT and MRI images. We denote the thick slices as $D_S = \{(x_s, y_s) | x_s \in R^{H \times W \times 3}, y_s \in R^{H \times W}\}$, and the thin-slice images as $D_T = \{x_t | x_t \in R^{H \times W \times 3}\}$. For the image extraction, we use the ResNet-34 that is pretrained on the ImageNet dataset as the encoder. For the decoder, we adopt the sub-pixel convolution to upsample and construct the images without heavy computations while bringing additional information to the prediction. In specific, the sub-pixel convolution can be derived as:

$$F^L = SP(W_L * F^{L-1} + b_L), \qquad (1)$$

where $SP(\cdot)$ operator arranges a tensor shaped in $H \times W \times C \times r^2$ into a tensor shaped in $rH \times rW \times C$. $F^L$ and $F^{L-1}$ are the feature maps of the image.

For training the model, we use the thick-slice images and thin-slice images as the inputs of the model, and optimize the model with the following objective function:

$$L(x_s, x_t) = L_S(p_s, y_s) + L_T(p_t), \tag{2}$$

where $p_s$ and $p_t$ are the model's predictions, and $y_s$ are the label of thick-slice images. More concrete, the $L_S$ is the cross-entropy loss which is computed as:

$$L_S = -\frac{1}{HW} \sum_{n=1}^{HW} \sum_{c=1}^{C} y_s^{n,c} \log p_t^{n,c}, \tag{3}$$

And $L_T$ is the entropy loss that guides the segmentation of the thin slices, which is obtained by:

$$L_T = -\frac{1}{C} \sum_{c=1}^{C} f(C p_t^{n,c}); \quad f(x) = x^2 - 1. \tag{4}$$

During model training, we iteratively optimize the above loss functions. For testing, we feed the image slices as the input into the trained model and get the predicted segmentation.

Compared to the previous medical segmentation approaches, they tend to train the networks from scratch within end-to-end manner. However, it has been widely proved that using the pre-trained model on the large-scale dataset, e.g., ImageNet can enable the network to learn the textual and shape prior effectively in the early stage of training. Therefore, we incorporated the ImageNet pretrained ResNet as the initialization of the encoder for obtaining powerful image representation during the first stage of training.

**2.7 Calculation of VV and ICV**

For this part, we present a method how to automatically calculate the VV and ICV using the segmentation method described in the previous sections. We first extract the ratio of the image from the scanned MRI or CT files, we denote this ratio $r = (r_x, r_y, r_z)$ (mm³ / pixel). Given an input MRI / CT volume, we first used the ventricle segmentation model to predict each type of ventricle, i.e., left lateral ventricle, right lateral ventricle, third ventricle, and fourth ventricle. For the i-th slice of the volume $V$, the prediction of each pixel is $V_{jk}^i$. Therefore, for the ventricle categorized as c, we calculate its $VV_c$ as follows:

$$VV_c = r_x r_y r_z \sum_i \sum_j \sum_k \mathbf{1}[V^i_{jk} = c], \tag{5}$$

where $\mathbf{1}[x = y]$ is the indicator function.

Therefore, given a scanned CT / MRI volume, we can use the above equation to estimate the VV of each ventricle. Similarly, we can estimate the volume of the whole brain by training a whole-brain segmentation model as we have stated in the previous section. Then, the volume of the brain $ICV$ can be calculated. We use the trained whole-brain segmentation model to predict the pixel. For the i-th slice of the volume $V$, the prediction of each pixel is $V^i_{jk}$. With the whole-brain region categorized as 1, we can calculate the corresponding ICV as follows:

$$ICV = r_x r_y r_z \sum_i \sum_j \sum_k \mathbf{1}[V^i_{jk} = 1], \tag{6}$$

Then, according to the definition, the VV/ICV can be obtained as follows:

$$\frac{VV}{ICV} = \frac{\sum_c VV_c}{ICV} \times 100\%. \tag{7}$$

**2.8 Statistical analysis**

Combining statistical methods used in previous literature [22, 30], we use Dice's similarity coefficient (DSC), Intraclass correlation coefficients (ICC), Pearson correlation, and Bland–Altman analysis to evaluate the spatial overlap, reliability, correlation, and consistency between the results of automatic and manual ventricle segmentation.

2.9 Implementation Detail

For the implementation, we first trained the model on the thick-slice datasets with SGD optimizer for 200 epochs. The initial learning rate was set to 1e-3 with linear decay schedule. Then, we used the pretrained model as the initialization of the model and applied both thick-slice and thin-slices on the proposed objective function with the initial learning rate 1e-4 for 100 epochs. The weight decay factor was 1e-5 for training. For the training time, the pre-train stage took about 5 hours on the machine with 4 NVIDIA TITAN RTX GPUs, and the main training took 10 hours on the same machine.

**3 Results**

**3.1 The processing results of CT image**

The DSC, ICC, and Pearson correlations of the VV generated by our model for automatic segmentation and the VV generated by manual segmentation are 0.95, 0.99, and 0.99, respectively. The DSC, ICC, and Pearson correlations of ICV generated by automatic segmentation and manual segmentation are 0.96, 0.99, and 0.99, respectively (Table 3 and Figure 2). Bland-Altman analysis shows that manual and automatic segmentation bias mean±standards deviations of VV and ICV are 4.2 ± 2.6 and 6.0 ± 3.8 (Figure 3). It takes 3.4 ± 0.3 seconds for our model to automatically segment the VV and ICV of a patient (Table 4).

### 3.2 The processing results of MRI image

The DSC, ICC, and Pearson correlations of the VV generated by automatic segmentation and manual segmentation are 0.94, 0.99, and 0.99, respectively. The DSC, ICC, and Pearson correlations of ICV generated by automatic segmentation and manual segmentation are 0.93, 0.99, and 0.99, respectively (Table 3 and Figure 2). Bland-Altman analysis shows that manual and automatic segmentation bias mean±standards deviations of VV and ICV are 2.0 ± 0.6 and 7.9 ± 3.8 (Figure 3). It takes 1.9 ± 0.1 seconds for our model to automatically segment the VV and ICV of a patient (Table 4).

### 3.3 Processing result display

For qualitatively evaluating the effectiveness of our proposed method, we visualize the segmentation results as well as their corresponding 3D reconstruction results for both MRI and CT samples in Figure 4. As depicted in the Figure, the right lateral ventricle is colored in red; the left lateral ventricle is colored in green; the yellow-colored region represents the third ventricle, and the blue region represents the fourth ventricle. Besides, the 3D segmentation results are visualized from the axial plane, the coronal plane, and the sagittal plane. We can notice that our method could not only predict the two ventricles mentioned above but also segment the third ventricle and the fourth ventricle well. By comparing the results of our method with other methods in the automatic segmentation of the ventricle of the validation set, we can see that our method is superior to the other two methods in automatically segmenting the ventricle (Table 5).

## 4 Discussions

The results produced by the multi-modal automatic ventricle segmentation method established through the brain images of NPH patients and the results produced by manual segmentation have excellent spatial overlap, reliability, correlation, and consistency. Therefore, the artificial intelligence method can realize the efficient and accurate assessment of the ventricular condition of NPH patients.

Volume measurement is obtained by segmentation, which refers to the process of describing the structure in imaging research [23]. The segmentation of the ventricle provides a quantitative measurement for the changes of the ventricle, forming important diagnostic information [31]. For manual segmentation, it took about 30 minutes to obtain the VV and ICV of a patient by manual segmentation [17]. There is no doubt that this will hinder the evaluation of the ventricles of large-scale samples [13]. Because of this, manual segmentation is often impractical in large-scale clinical practice, and more automated methods are urgently needed to complete it [32, 33]. So, the automated brain image segmentation method is a research hotspot in recent years [16]. More importantly, The development of accurate, fast, and easy-to-use ventricular volume segmentation methods is of great significance for further research and evaluation of the standardized use of ventricular volume in NPH patients [27]. The automatic ventricle segmentation method can overcome the limitations of the manual segmentation method [10]. It is an efficient and rapid ventricle segmentation method [12, 23]. The most important thing is that it can significantly shorten the operation time [24]; this lays a solid foundation for the direct measurement of the ventricle volume in large-scale clinical practice. However, due to the difference between CT and MRI images, automatic ventricle segmentation methods based on CT images are often difficult to process MR images [34]. On the other hand, the existing automatic ventricle segmentation methods for NPH patients only segment the ventricle structure [24]. Compared with the volume of the ventricle, the VV/ICV can reflect the situation of the ventricle in the whole skull [22] and take into account the differences in the volume of each subject caused by changes in anatomy [35].

Compared with current semi-supervised segmentation techniques, Li et al. [36] proposed a generative method that adopts StyleGAN2 with an augmented label synthesis branch, which utilizes partially labeled images to predict the out-of-domain images. During the inference time, it requires to finetune the network to find the best reconstruction with their proposed objective function for the target image. By contrast, our method finetunes the network with labeled and unlabeled images together to preserve the generalization ability for both thick-slice and thin-slice images under both CT and MRI modalities.

This study has the following limitations: first of all, this is a retrospective study; it is based on existing images and clinical information. Next, we will collect more comprehensive clinical information and imaging data of NPH patients through prospective research; we will also pay more attention to the follow-up process of NPH patients and the situation after surgery; this will help us to understand NPH disease more comprehensively. Secondly, our study is a single-center study; patients are all from a single area. Previous studies mentioned that the ventricles of different races are different [17]. Then, we will conduct multi-center research to further understand the ventricles of NPH patients and improve the applicability and application value of the automatic ventricle segmentation method. Third, we only performed routine brain imaging analysis. Functional imaging is also very important for the diagnosis and treatment of NPH patients [37]. Therefore, in follow-up research, we will use conventional imaging and functional imaging to further understand the changes in brain structure and function of NPH patients. The last but most important thing is that both the medical field and the artificial intelligence field are constantly evolving. Our approach needs to finetune the network on the thin-slice and thick-slice together for preserving the generalization ability on both types of images, which requires to access the thick-slice images when training for the thin-slice images although it works for multiple modalities. This would be impractical when it comes to the privacy of the data. Subsequently, we will continue to optimize the algorithm of the model to meet the actual needs of clinical practice. To better integrate artificial intelligence and medicine, complement each other and make progress together.

## 5 Conclusion

In summary, we have established a multi-modal and high-performance automatic ventricle segmentation method to achieve efficient and accurate automatic measurement of the ventricular volume of NPH patients. It can not only process CT and MRI images at the same time but also calculate the ventricular volume and relative ventricular volume at the same time. The whole process is relatively fast compared to the traditional method. Besides, the thickness agnostic ventricle and whole-brain segmentation can handle the samples generated by different scanners as well as the thickness of slices. This is an effective combination of the medical field and the field of artificial intelligence. It not only lays a solid foundation for the subsequent analysis of large samples of NPH patients, but also helps clinicians quickly and accurately understand the situation of normal pressure hydrocephalus patient's ventricles, which can help clinicians in the diagnosis of NPH patients, the follow-up process, and the evaluation of surgical effects.

**Table legends**

**Table 1** Characteristics data of the definite NPH patients.

All results are in the form of mean ± standard deviation

MMSE = Mini-mental State Examination; TUG = Timed Up and Go Test; iNPHGS = iNPH Grading Scale

**Table 2** Scan parameters of CT and MRI.

*: The device does not have this parameter

TR = Repetition Time; TE = Echo Time

**Table 3** The DSC, ICC, and Pearson correlation of validation set manual and automatic measurement results.

All results are in the form of mean ± standard deviation

ICC = Intraclass Correlation Coefficient; DSC = Dice Similarity Coefficient

**Table 4** The validation set measurement results.

All results are in the form of mean ± standard deviation

**Table 5** Comparison results (DSC) of our method with other methods for automatically segmenting the ventricle of the validation set.

All results are in the form of mean ± standard deviation

DSC = Dice similarity coefficient

**Figure legends**

**Fig. 1** Flow chart and network structure of this research.

**Fig. 2** Pearson correlation analysis diagram of manual and automatic segmentation results. Whether it is the Ventricle Volume (VV) and Intracranial Volume (ICV) of the CT image or MRI image; the Pearson correlation between automatic and manual segmentation results is 0.99, and there is the statistical significance ($P<0.01$).

**Fig. 3** Bland-Altman analysis diagram of manual and automatic segmentation results. In the CT image, the Bland - Altman analysis shows that manual and automatic segmentation bias mean ± standards deviations of VV and ICV are $4.2 \pm 2.6$ and $6.0 \pm 3.8$. In the MRI image, the Bland - Altman analysis shows that manual and automatic segmentation bias mean±standards deviations of VV and ICV are $2.0 \pm 0.6$ and $7.9 \pm 3.8$.

**Fig. 4** The visualization of the 3D brain ventricles and whole-brain segmentation results with two different modalities (CT / MRI) and the corresponding three-dimensional visualization of the predictions. The right lateral ventricle is colored in red; the left lateral ventricle is the green one; the third ventricle is colored with yellow, and the blue region indicates the fourth ventricle.

**Acknowledgments**

The author is very grateful to Mengyao Xu and Yibo Xu for their guidance on the experimental design of this study, Chuming Xu and Xiaolian Li for their sincere help in data collection, Weiwen Zhou and Fengping Huang for their suggestion on data analysis. This study is supported in part by Project of Shenzhen International Cooperation Foundation (GJHZ20180926165402083), in part by the funding support from the Basque Government (Eusko Jaurlaritza) through the Consolidated Research Group MATHMODE (IT1294-19), in part by the British Heart Foundation (Project Number: TG/18/5/34111, PG/16/78/32402), in part by the Hangzhou Economic and Technological Development Area Strategical Grant (Imperial Institute of Advanced Technology), in part by the European Research Council Innovative Medicines Initiative on Development of Therapeutics and Diagnostics Combatting Coronavirus Infections Award "DRAGON: rapiD and secuRe AI imaging based diaGnosis, stratification, fOllow-up, and preparedness for coronavirus paNdemics" [H2020-JTI-IMI2 101005122], in part by the AI for Health Imaging Award "CHAIMELEON: Accelerating the Lab to Market Transition of AI Tools for Cancer Management" [H2020-SC1-FA-DTS-2019-1 952172], and in part by the UK Research and Innovation Future Leaders Fellowship [MR/V023799/1].

**Conflicts of interest/Competing interests**

There are no potential competing interests in our paper. And all authors have seen the manuscript and approved to submit to your journal. We confirm that the content of the manuscript has not been published or submitted for publication elsewhere.

**Availability of data and material**

All datasets analyzed during the present study are available from the corresponding author on reasonable request.

**Code availability**

Opensource codes will be uploaded and published on Github subject to the publication of this work.

**Author's contributions**

XZ, QY, XY, JC, HM, JX, JS, and GY contributed to study design and writing – original draft preparation. XZ, XY, JC, HM, and JX contributed to data collection. QY, JX, JS and GY contributed to data visualization. XZ, QY, JX, JS, and GY contributed to writing – review and editing. JX, JS, and GY contributed to supervision and funding acquisition.

**Ethics approval**

This study passed the ethical approval of The First Affiliated Hospital of Shenzhen University's bioethics committee (approval no. KS20190114001), and the researchers all signed the informed consent form.

**Consent to participate**

Informed consent was obtained from all individual participants included in the study.

**Consent for publication**

All authors have read and agreed to the published version of the manuscript.

**References:**

1. Adams RD, Fisher CM, Hakim S, Ojemann RG, Sweet WH (1965) Symptomatic occult hydrocephalus with normal cerebrospinal-fluid pressure: a treatable syndrome. N Engl J Med 273:117-126. https://doi.org/10.1056/NEJM196507152730301

2. Nakajima M, Yamada S, Miyajima M (2021) Guidelines for Management of Idiopathic Normal Pressure Hydrocephalus (Third Edition): Endorsed by the Japanese Society of Normal Pressure Hydrocephalus. Neurol Med Chir (Tokyo) 61:63-97. https://doi.org/10.2176/nmc.st.2020-0292

3. He W, Fang X, Wang X (2020) A new index for assessing cerebral ventricular volume in idiopathic normal-pressure hydrocephalus: a comparison with Evans' index. Neuroradiology 62:661-667. https://doi.org/10.1007/s00234-020-02361-8

4. Prince M, Bryce R, Albanese E, Wimo A, Ribeiro W, Ferri CP (2013) The global prevalence of dementia: a systematic review and metaanalysis. Alzheimers & Dementia 9:63-75. https://doi.org/10.1016/j.jalz.2012.11.007

5. Jaraj D, Rabiei K, Marlow T, Jensen C, Skoog I, Wikkelso C (2014) Prevalence of idiopathic normal-pressure hydrocephalus. Neurology 82:1449-1454. https://doi.org/10.1212/WNL.0000000000000342

6. Kazui H, Miyajima M, Mori E, Ishikawa M (2015) Lumboperitoneal shunt surgery for idiopathic normal pressure hydrocephalus (SINPHONI-2): an open-label randomised trial. Lancet Neurology 14:585-594. https://doi.org/10.1016/S1474-4422(15)00046-0

7. Andren K, Wikkelso C, Tisell M, Hellstrom P (2014) Natural course of idiopathic normal pressure hydrocephalus. J Neurol Neurosurg Psychiatry 85:806-810. https://doi.org/10.1136/jnnp-2013-306117

8. Jaraj D, Wikkelso C, Rabiei K (2017) Mortality and risk of dementia in normal-pressure hydrocephalus: A population study. Alzheimers & Dementia 13:850-857. https://doi.org/10.1016/j.jalz.2017.01.013

9. Palm WM, Saczynski JS, van der Grond J (2009) Ventricular dilation: association with gait and cognition. Annals of Neurology 66:485-493. https://doi.org/10.1002/ana.21739

10. Kocaman H, Acer N, Köseoğlu E, Gültekin M, Dönmez H (2019) Evaluation of intracerebral ventricles volume of patients with Parkinson's disease using the atlas-based method: A methodological study. Journal of Chemical Neuroanatomy 98:124-130. https://doi.org/10.1016/j.jchemneu.2019.04.005

11. Kempton MJ, Underwood TSA, Brunton S (2011) A comprehensive testing protocol for MRI neuroanatomical segmentation techniques: Evaluation of a novel lateral ventricle segmentation method. Neuroimage 58:1051-1059. https://doi.org/10.1016/j.neuroimage.2011.06.080

12. Quon JL, Han M, Kim LH (2021) Artificial intelligence for automatic cerebral ventricle segmentation and volume calculation: a clinical tool for the evaluation of pediatric hydrocephalus. Journal of Neurosurgery: Pediatrics 27:131-138. https://doi.org/10.3171/2020.6.PEDS20251

13. Dubost F, Bruijne MD, Nardin M (2020) Multi-atlas image registration of clinical data with automated quality assessment using ventricle segmentation. Medical Image Analysis 63:101698. https://doi.org/10.1016/j.media.2020.101698

14. Qiu W, Yuan J, Rajchl M (2015) 3D MR ventricle segmentation in pre-term infants with post-hemorrhagic ventricle dilatation (PHVD) using multi-phase geodesic level-sets. Neuroimage 118:13-25. https://doi.org/10.1016/j.neuroimage.2015.05.099

15 Poh LE, Gupta V, Johnson A, Kazmierski R, Nowinski WL (2012) Automatic Segmentation of Ventricular Cerebrospinal Fluid from Ischemic Stroke CT Images. Neuroinformatics 10:159-172. https://doi.org/10.1007/s12021-011-9135-9

16 Cherukuri V, Ssenyonga P, Warf BC, Kulkarni AV, Monga V, Schiff SJ (2018) Learning Based Segmentation of CT Brain Images: Application to Postoperative Hydrocephalic Scans. IEEE Trans Biomed Eng 65:1871-1884. https://doi.org/10.1109/TBME.2017.2783305

17 Ambarki K, Israelsson H, Wåhlin A, Birgander R, Eklund A, Malm J (2010) Brain Ventricular Size in Healthy Elderly. Neurosurgery 67:94-99. https://doi.org/10.1227/01.NEU.0000370939.30003.D1

18 Hassan MM, Alam MGR, Uddin MZ, Huda S, Almogren A, Fortino G (2019) Human emotion recognition using deep belief network architecture. Information Fusion 51:10-18. https://doi.org/10.1016/j.inffus.2018.10.009

19 Zhang Y, Gravina R, Lu H, Villari M, Fortino G (2018) PEA: Parallel electrocardiogram-based authentication for smart healthcare systems. Journal of Network and Computer Applications 117:10-16. https://doi.org/10.1016/j.jnca.2018.05.007

20 Piccialli F, Somma VD, Giampaolo F, Cuomo S, Fortino G (2021) A survey on deep learning in medicine: Why, how and when? Information Fusion 66:113-137. https://doi.org/10.1016/j.inffus.2020.09.006

21 Yang G, Ye Q, Xia J (2022) Unbox the black-box for the medical explainable AI via multi-modal and multi-centre data fusion: A mini-review, two showcases and beyond. Information Fusion 77: 29-52. https://doi.org/10.1016/j.inffus.2021.07.016

22 Ntiri EE, Holmes MF, Forooshani PM (2021) Improved Segmentation of the Intracranial and Ventricular Volumes in Populations with Cerebrovascular Lesions and Atrophy Using 3D CNNs. Neuroinformatics. https://doi.org/10.1007/s12021-021-09510-1

23 Huff TJ, Ludwig PE, Salazar D, Cramer JA (2019) Fully automated intracranial ventricle segmentation on CT with 2D regional convolutional neural network to estimate ventricular volume. International Journal of Computer Assisted Radiology and Surgery 14:1923-1932. https://doi.org/10.1007/s11548-019-02038-5

24 Shao M, Han S, Carass A (2019) Brain ventricle parcellation using a deep neural network: Application to patients with ventriculomegaly. Neuroimage Clin 23:101871. https://doi.org/10.1016/j.nicl.2019.101871

25 Zhao S, Gao Z, Zhang H et al (2017) Robust Segmentation of Intima-Media Borders with Different Morphologies and Dynamics During the Cardiac Cycle. IEEE Journal of Biomedical and Health Informatics 22:1571-1582. https://doi.org/10.1109/JBHI.2017.2776246

26 Zhao S, Wu X, Chen B, Li S (2021) Automatic vertebrae recognition from arbitrary spine MRI images by a category-Consistent self-calibration detection framework. Medical Image Analysis 67:101826. https://doi.org/10.1016/j.media.2020.101826

27 Neikter J, Agerskov S, Hellström P (2020) Ventricular Volume Is More Strongly Associated with Clinical Improvement Than the Evans Index after Shunting in Idiopathic Normal Pressure Hydrocephalus. American Journal of Neuroradiology 41:1187-1192. https://doi.org/10.3174/ajnr.A6620

28 Mori E, Ishikawa M, Kato T (2012) Guidelines for management of idiopathic normal pressure hydrocephalus: second edition. Neurol Med Chir (Tokyo) 52:775-809. https://doi.org/10.2176/nmc.52.775

29 Zhou X, Ye Q, Jiang Y (2020) Systematic and Comprehensive Automated Ventricle Segmentation on Ventricle Images of the Elderly Patients: A Retrospective Study. Frontiers in Aging Neuroscience 12:618538. https://doi.org/10.3389/fnagi.2020.618538

30 Zhao SX, Xiao YH, Lv FR, Zhang ZW, Sheng B, Ma HL (2018) Lateral ventricular volume measurement by 3D MR hydrography in fetal ventriculomegaly and normal lateral ventricles. Journal of Magnetic Resonance Imaging 48:266-273. https://doi.org/10.1002/jmri.25927

31 Chen W, Smith R, Ji S, Ward KR, Najarian K (2009) Automated ventricular systems segmentation in brain CT images by combining low-level segmentation and high-level template matching. BMC Medical Informatics and Decision Making 9:S4. https://doi.org/10.1186/1472-6947-9-S1-S4

32 Chou Y, Leporé N, de Zubicaray GI (2008) Automated ventricular mapping with multi-atlas fluid image alignment reveals genetic effects in Alzheimer's disease. Neuroimage 40:615-630. https://doi.org/10.1016/j.neuroimage.2007.11.047

33 Tang X, Crocetti D, Kutten K (2015) Segmentation of brain magnetic resonance images based on multi-atlas likelihood fusion: testing using data with a broad range of anatomical and photometric profiles. Frontiers in Neuroscience 9:61. https://doi.org/10.3389/fnins.2015.00061

34 Qian X, Lin Y, Zhao Y, Yue X, Lu B, Wang J (2017) Objective Ventricle Segmentation in Brain CT with Ischemic Stroke Based on Anatomical Knowledge. Biomed Research International 2017:1-11. https://doi.org/10.1155/2017/8690892

35 Tarnaris A, Toma AK, Pullen E (2011) Cognitive, biochemical, and imaging profile of patients suffering from idiopathic normal pressure hydrocephalus. Alzheimers & Dementia 7:501-508. https://doi.org/10.1016/j.jalz.2011.01.003

36 Li D, Yang J, Kreis K, Torralba A, Fidler S (2021) Semantic segmentation with generative models: Semi-supervised learning and strong out-of-domain generalization. Proceedings of the IEEE/CVF Conference on Computer Vision and Pattern Recognition 2021: 8300-8311.

37 Zhang H, He WJ, Liang LH (2021) Diffusion Spectrum Imaging of Corticospinal Tracts in Idiopathic Normal Pressure Hydrocephalus. Frontiers in Neurology 12:636518. https://doi.org/10.3389/fneur.2021.636518

**Table 1. Characteristics data of the definite NPH patients**

| Characteristics | Value |
|---|---|
| **Age(years)** | 72.3 ± 6.9 |
| **Sex(male/female)** | 79/64 |
| **MMSE score** | 21.8 ± 4.7 |
| **TUG (second)** | 27.6 ± 18.3 |
| **iNPHGS score** | 5.8 ± 2.4 |

All results are in the form of mean ± standard deviation.

MMSE=Mini-mental State Examination; TUG=Timed Up and Go Test; iNPHGS=iNPH Grading Scale.

**Table 2. Scan parameters of CT and MRI**

| Name | Type | Producer | Field Strength (T) | Sequence | TR (ms) | TE (ms) | Flip Angle (°) | Slice Thickness (mm) | Pixel | Examination Quantity |
|---|---|---|---|---|---|---|---|---|---|---|
| **A** | MRI | General Electric | 1.5 | T1 | 1910 | 23.6 | 90 | 6 | 0.4688*0.4688*8 | 45 |
| **C** | MRI | Siemens | 1.5 | T1 | 388 | 13 | 90 | 6 | 0.6875*0.6875*7.8 | 56 |
| **D** | MRI | Siemens | 3.0 | T1 | 2000 | 7.4 | 150 | 6 | 0.6875*0.6875*7.8 | 49 |
| **E** | CT | Siemens | NA# | NA# | NA# | NA# | NA# | 5 | 0.3906*0.3906*5 | 59 |
| **F** | CT | Siemens | NA# | NA# | NA# | NA# | NA# | 4.8 | 0.4199*0.4199*4.825 | 63 |

#: The device does not have this parameter.

TR=Repetition Time; TE=Echo Time

**Table 3. The DSC, ICC and Pearson correlation of validation set manual and automatic measurement results**

| Manual and automatic measurement results | CT | | | MRI | | |
|---|---|---|---|---|---|---|
| | DSC | ICC | Pearson Correlation | DSC | ICC | Pearson Correlation |
| **Ventricle Volume** | 0.95 ± 0.01 | 0.99 | 0.99 | 0.94 ± 0.01 | 0.99 | 0.99 |
| **Intracranial volume** | 0.96 ± 0.02 | 0.99 | 0.99 | 0.93 ± 0.03 | 0.99 | 0.99 |

All results are in the form of mean ± standard deviation.

ICC= Intraclass Correlation Coefficient; DSC= Dice similarity coefficient

**Table 4** Validation set measurement results

| | CT | | MRI | |
|---|---|---|---|---|
| | Manual Segmentation | Auto Segmentation | Manual Segmentation | Auto Segmentation |
| **Ventricular volume(ml)** | 136.6 ± 34.6 | 132.4 ± 35.5 | 124.6 ± 27.6 | 122.6 ± 27.5 |
| **Intracranial volume(ml)** | 1396.1 ± 144.6 | 1390.2 ± 144.4 | 1238.7 ± 112.1 | 1231.6 ± 111.4 |
| **Time consuming(s)** | >1000 | 3.4 ± 0.3 | >1000 | 1.9 ± 0.1 |

All results are in the form of mean ± standard deviation

**Table 5** Comparison results (DSC) of our method with other methods for automatically segmenting the ventricle of the validation set

| Methods | CT | | MRI | |
| --- | --- | --- | --- | --- |
| | Ventricle Volume | Intracranial volume | Ventricle Volume | Intracranial volume |
| **Ours** | 0.95 ± 0.01 | 0.96 ± 0.02 | 0.94 ± 0.01 | 0.93 ± 0.03 |
| **U-Net** | 0.90 ± 0.03 | 0.88 ± 0.02 | 0.89 ± 0.03 | 0.87 ± 0.02 |
| **U-Net++** | 0.90 ± 0.02 | 0.89 ± 0.01 | 0.90 ± 0.02 | 0.90 ± 0.03 |

All results are in the form of mean ± standard deviation

DSC = Dice similarity coefficient

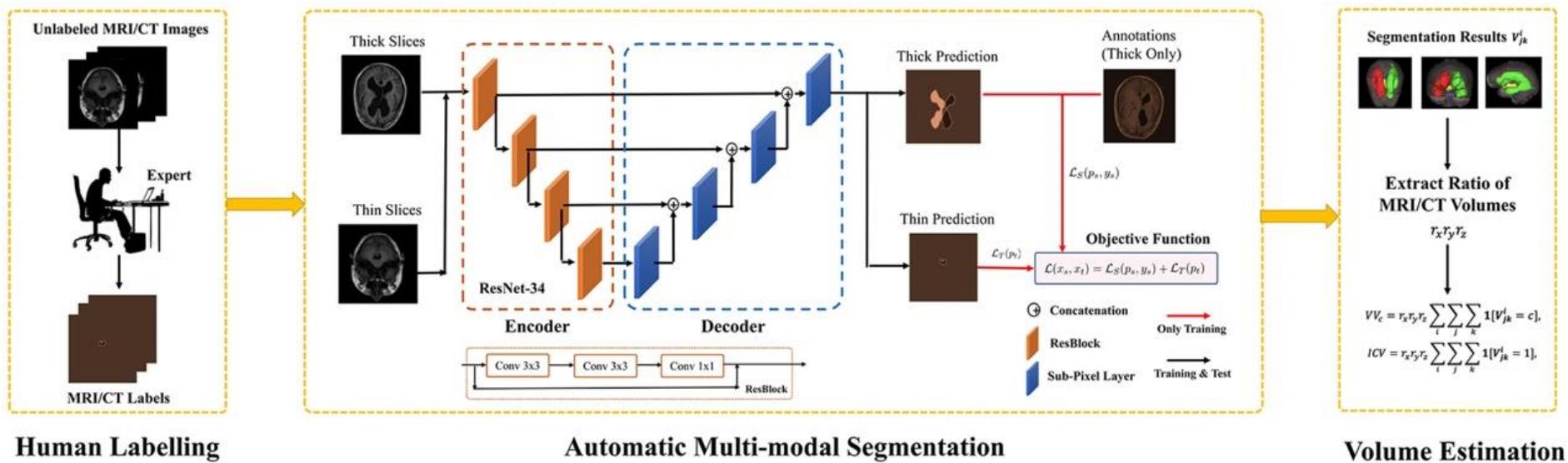

Figure 1

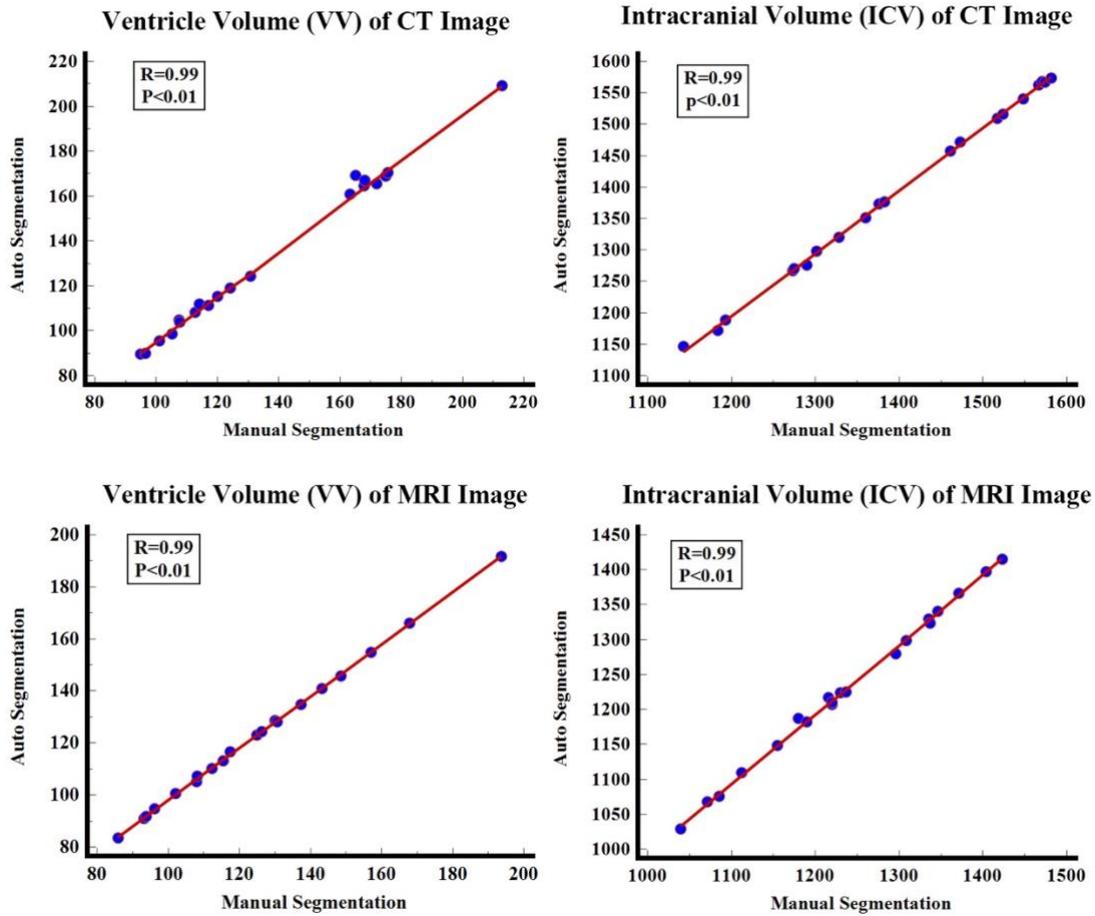

Figure 2

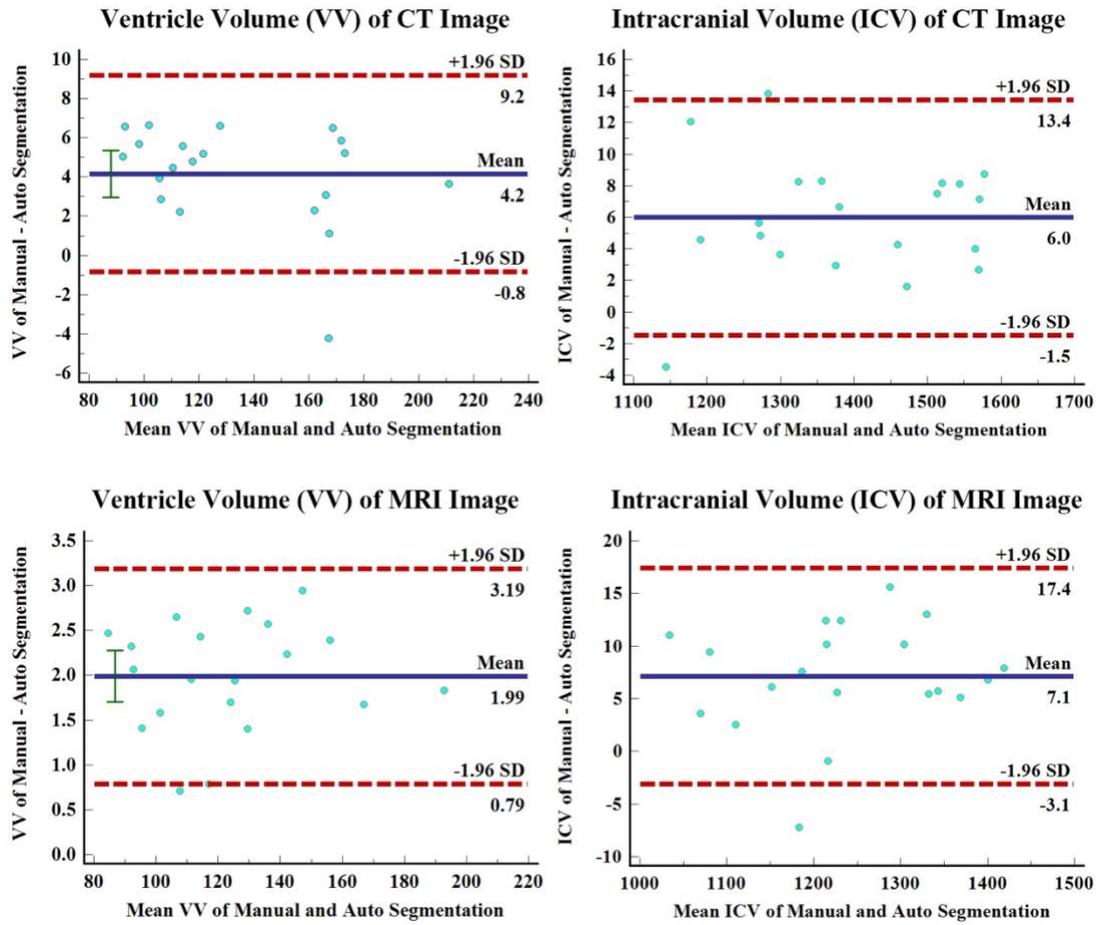

Figure 3

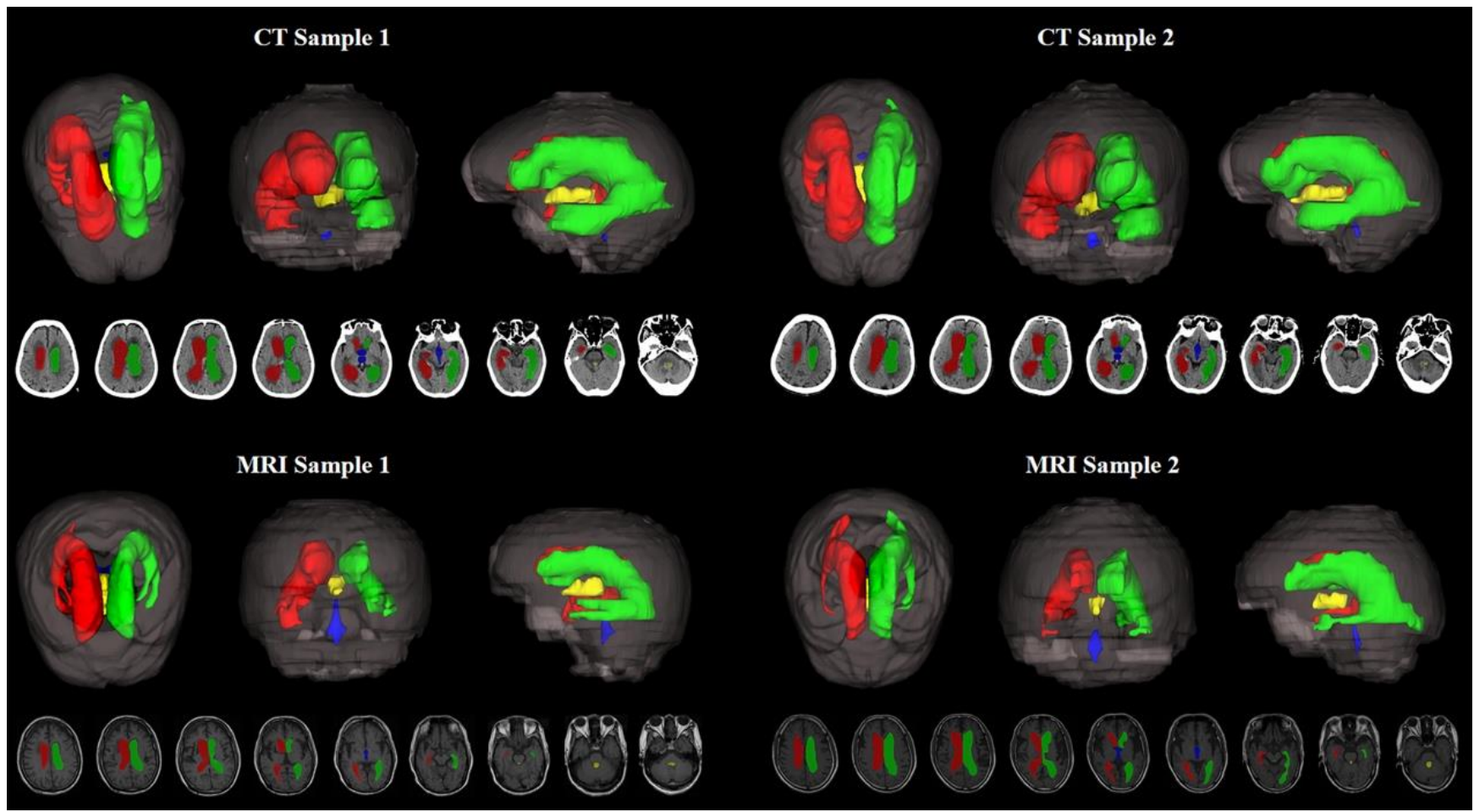

Figure 4